\documentclass[10pt]{article}
\usepackage{axodraw}
\begin{document}
\begin{flushright}
\texttt{hep-ph/0309110}\\
SINP/TNP/03-31\\
\end{flushright}

\vskip 50pt

\begin{center}
{\Large \bf Ultraviolet sensitivity of rare decays in 
nonuniversal \\ extra dimensional models} \\
\vspace*{1cm}
\renewcommand{\thefootnote}{\fnsymbol{footnote}}
{\large {\sf Paramita Dey} and {\sf Gautam Bhattacharyya} 
}
\vskip 5pt
{ \em Saha Institute of Nuclear Physics, 1/AF Bidhan Nagar,
 Kolkata 700064, India}\\
\normalsize
\end{center}

\begin{abstract}

We consider a nonuniversal five dimensional model in which fermions
are localised on a four dimensional brane, while gauge bosons and a
scalar doublet can travel in the bulk. As a result of KK number
non-conservation at the brane-bulk intersection, the ultraviolet
divergence does not cancel out in some physical observables.  For
example, the $B_d \to \ell^+ \ell^-$ decay amplitude is linearly
divergent, while $B$--$\bar{B}$ mixing amplitude is log divergent. We
attempt to identify the exact source of this nonrenormalizability. We
compare and contrast our results with those obtained in the universal
five dimensional model where all particles travel in the extra
dimension.

\vskip 5pt \noindent
\texttt{PACS Nos:~11.10.Kk, 12.60.-i} \\
\texttt{Key Words:~ Nonuniversal extra dimension, Kaluza-Klein tower,
$B$ decay}
\end{abstract}

\renewcommand{\thesection}{\Roman{section}}
\setcounter{footnote}{0}
\renewcommand{\thefootnote}{\arabic{footnote}}

\section{Introduction}
The idea that extra compactified space-like dimension may exist has
caught the imagination of physicists working in phenomenology, string
theory and cosmology alike. This had led to an important cross-road
where perspectives from all the three avenues could melt, and as a
result recent years have seen a huge boom of activity in this
area. The possibility that the size of the extra dimension may be
large \cite{add}, much higher than the Planck length of $\sim
10^{-33}$ cm, perhaps to the tune of ${\rm TeV}^{-1} \sim 10^{-18}$ cm
to be accessed by the standard model (SM) fields \cite{anto}, or even
as large as a millimeter where gravity or sterile neutrino could
propagate \cite{ddg,addj,smirnov,pilaft,gb}, has brought the theory
within the testable domain of the ongoing and future experiments.

In this paper, we consider a TeV scale extra dimensional model in
which the SM fermions are four dimensional fields while the bosons can
travel in the higher dimensional bulk \cite{pq}.  Operationally, a
higher dimensional theory can be described as an effective theory in
four dimension which contains an infinite tower of four dimensional
Kaluza-Klein (KK) modes of the higher dimensional fields. As a result,
such effective theories are nonrenormalizable. In the conventional
nonrenormalizable theories, at least at the tree level all processes
are finite. But when we have more than one extra dimension, even the
tree amplitude diverges. The theory then depends on two additional
parameters -- the compactification scale ($R^{-1} = M_c$) and the
ultraviolet cut-off ($M_s = n_s/ R$). For one extra dimension the tree
amplitude is finite. But higher order processes are, in general,
cut-off dependent.

In theories with only one extra dimension, another important
observation \cite{santa1} is that if the one loop diagrams involve a
{\em single} summation over the KK modes then the theory is as well
behaved as its zero mode counterpart. Then the total amplitude depends
only on $M_c$ and {\em not} on $M_s$. On the contrary, if there is
more than one independent sum over KK modes inside a one loop
integral, the total amplitude diverges.

The localisation of fermions on a four dimensional brane was motivated
in \cite{pq} from a stringy perspective that chiral matters should be
placed in the twisted sector while non-chiral states can travel in the
bulk. We refer to this picture as the nonuniversal extra dimensional
(NUED) model.  In the present analysis we deal with a NUED model
having only one extra dimension which is accessed only by the gauge
bosons and the scalar doublet.  Constraints on this class of models
from electroweak observables were placed in \cite{ew}.  The
implications of this scenario in the context of $Z\to b\bar{b}$ decay
and Kaon and $B$ meson mixings have been studied in \cite{santa1}.

On the other hand, one may consider a situation in which all SM
particles are allowed to travel in all available extra dimensions.
These are called universal extra dimensional (UED) models. The
simplest of this type, with only one extra dimension, was constructed
and its implications to oblique electroweak parameters were studied in
\cite{acd}. A detailed phenomenological implications, mainly related
to FCNC processes, have been investigated in
\cite{buras1,buras2,debrupa}. Implications of the UED scenario to
$Z\to b\bar{b}$ decay has been examined in \cite{santa2}. Other
phenomenological implications in different variants of UED scenario
have been studied in Refs.~\cite{uedothers}.

The core issue that distinguishes the ultraviolet behaviour of the UED
and NUED models is the question of KK momentum
conservation. Corresponding to every noncompact spatial dimension
there is a conserved momentum. When that direction is compactified the
momentum becomes discrete ($n/R$) but still remains a conserved
quantity. The situation becomes tricky in the NUED models when, for
example, a boson in the bulk couples to two fermions localised at the
brane. The momentum in the extra dimension is not then conserved as a
consequence of the breakdown of translational invariance at the
brane. In other words, KK number does not remain conserved at the
brane-localised interaction vertex. Any bulk interaction, however,
does conserve KK parity.  On the other hand, in the UED models all
interactions do conserve KK number. In a loop diagram involving KK
modes in the internal lines, the occurrence of a single summation or a
multiple summation can be linked to the issue of KK number
conservation or nonconservation. This constitutes the prime criterion
to judge whether the theory would be well-behaved or ultraviolet
sensitive.  The one loop corrections in the UED model with {\em only
one} extra dimension lead to finite results with $R$ as the new
physics parameter\footnote{Any order loop corrections in UED models
with more than one extra dimension are sensitive to the ultraviolet
cut-off. The exact nature of the cut-off dependence depends on the
number of extra dimensions \cite{acd}. Also, the degree of divergence
may be different in different processes. The same is true in NUED
models, only that the degree of divergence is higher.}.  On the
contrary, one loop corrections to a class of physical observables in
NUED models with one extra dimension are vulnerable against unknown
high scale physics, i.e., the results depend on $n_s$ as well. The
exact dependence indeed depends on the process concerned. When we go
to higher loops, UED models also give rise to divergent results for
physical observables. Indeed, the degree of divergence in NUED models
is higher than that in UED models at any order.

In this context, we concentrate on flavour changing neutral current
(FCNC) processes, in particular, the decay $B_d \to
\ell^+\ell^-$. Such a decay can proceed through penguin and box graphs
at leading order. The branching ratio is also GIM suppressed. We
investigate how the SM penguin and box graphs are modified due to the
presence of KK towers in the NUED framework. More specifically, we
probe which are the specific diagrams and interaction vertices that
are at the root of the ultraviolet sensitivity. We also revisit the
$B$--$\bar{B}$ mixing and $Z \to b\bar{b}$ decay in the context of
NUED scenario and agree with the results of \cite{santa1}. We finally
make a comment on the $t \to cZ$, $b \to s\gamma$, and $b \to s
\ell^+\ell^-$ decays in such a scenario.  We compare and contrast our
results with those obtained in the UED scenario
\cite{buras1,buras2,debrupa}.

\section{The basic features of the NUED model}
Here we set up our notations and recall the essential features of the
NUED model.

1. The extra dimension ($y$) is compactified on a circle of radius $R
   = M_c^{-1}$ and $y$ is identified with $-y$, i.e., it corresponds to
   an orbifold $S^1/Z_2$.

2. The fermions are four dimensional (4D) fields. They are all
   localised at the orbifold fixed point ($y = 0$).

3. The gauge bosons $A^M(x,y)$ and the scalar doublet $\phi(x,y)$
   are five dimensional (5D) fields. The 5D fields can be Fourier
   expanded in terms of 4D KK fields as  
\begin{eqnarray}
\label{fourier}
A^{\mu}(x,y)&=&\frac{1}{\sqrt{2\pi R}}A^{\mu}_{(0)}(x)+\frac{1}{\sqrt{\pi
R}}\sum^{\infty}_{n=1}A^{\mu}_{(n)}(x)\cos\frac{ny}{R},~~~~
A^5(x,y) = \frac{1}{\sqrt{\pi
R}}\sum^{\infty}_{n=1}A^5_{(n)}(x)\sin\frac{ny}{R}, \\
\phi^+(x,y)&=&\frac{1}{\sqrt{2\pi R}}\phi^+_{(0)}(x)+\frac{1}{\sqrt{\pi
R}}\sum^{\infty}_{n=1}\phi^+_{(n)}(x)\cos\frac{ny}{R},~~~~
\phi^-(x,y) = \frac{1}{\sqrt{\pi
R}}\sum^{\infty}_{n=1}\phi^-_{(n)}(x)\sin\frac{ny}{R}. 
\end{eqnarray}
Above, $x\equiv x^{\mu}$ ($\mu$=0,1,2,3 denote the four non-compact
space-time coordinates), $y$ denotes the fifth (extra) compactified
coordinate, and $M=$0,1,2,3,5. The fields $\phi^{\pm}_{(n)}(x)$ are
the 4D KK scalar fields, $A^{\mu}_{(n)}(x)$ are the 4D KK gauge fields
and $A^{5}_{(n)}(x)$ are the 4D KK scalar fields in the adjoint
representation of the gauge group. The field $A^5(x,y)$ depends on
sine of $y$ to ensure its absence on the brane ($y=0$).

4. A typical conventional bosonic 4D propagator is modified by the
presence of KK towers in the following way:
\begin{eqnarray} 
{(k_E^2+M^2)}^{-1} \longrightarrow
\sum_{n=-\infty}^{\infty} {(k_E^2+M^2+{n^2}/{R^2})}^{-1}
= (\pi R/k'_E) \coth (\pi R k'_E),
\end{eqnarray} 
where $k_E$ is an Euclidean four momentum and
$k'_E=\sqrt{k_E^2+M^2}$. For large argument hyperbolic cotangent
function goes like unity. Therefore, each modified propagator
increases the degree of divergence by reducing one power of $k_E$ in
the denominator. This means that if a conventional diagram is log
divergent in the ultraviolet limit, then KK towering for only one
propagator makes it linearly divergent. If two propagators are
modified, and the KK number for one is independent of the KK number of
the other, i.e., the two sums can be independently carried out, then a
logarithmically divergent diagram becomes quadratically divergent, and
so on. However, if by virtue of KK number conservation, the KK indices
of the two propagators are not independent, then the divergence will
be less than quadratic.

5. In principle, the summation over KK number cannot go up to
infinity. There should be an ultraviolet cutoff, which corresponds to
the scale $M_s$ at which unknown physics creeps in.

6. The KK number is not conserved when the interaction vertex is
   located at the brane, while any bulk interaction does conserve KK
   number.

Thus the parameters of the NUED model are $R$ and $M_s$. We trade them for
$a~(\equiv \pi R M_W)$ and $n_s~(\equiv M_s R)$.

\section{The decay \boldmath{$B_d \rightarrow\ell^+\ell^-$} in the SM}
In the SM, the contribution to the $B_d \to \ell^+\ell^-$ decay comes
from the penguin and box graphs. We work in the 't Hooft - Feynman
gauge and we denote the would-be Goldstone bosons $\phi^\pm$
separately from the gauge bosons $W^\pm$. First, we consider the
penguin part. The effective $Z\bar{b}d$ vertex is given by \cite{bb1}
\begin{eqnarray}
\label{smevz}
\Gamma_\mu^{Zbd} = -\frac{ig}{\cos\theta_W}\bigg(\frac{g^2}{16\pi^2}
F^{\rm SM}_{P}\xi_t\bigg)\gamma_\mu P_L,
\end{eqnarray}
where $\xi_t = V_{tb} V_{td}^*$, ~and ($x_t \equiv m_t^2/M_W^2$) 
\begin{eqnarray}
\label{smpf}
F^{\rm SM}_{P}=
\frac{x_t}{4}\Big[\frac{x_t-6}{x_t-1}+\frac{3x_t+2}{(x_t-1)^2}\ln
x_t\Big].
\end{eqnarray}
The penguin part of the decay amplitude is given by 
\begin{eqnarray}
\label{smpa}
\mathcal{M}^{\rm SM}_{P} = \frac{g^4}{64\pi^2 M_W^2} \xi_t 
\left(\frac{F^{\rm SM}_P}{2}\right) 
2t_{3{\ell}} ~(\bar\ell\ell)_{V-A} (\bar b d)_{V-A}. 
\end{eqnarray}

Next we include the box contribution. This is dominated by $W^\pm$
exchanged graphs. The amplitude is given by 
\begin{eqnarray}
\label{smba}
\mathcal{M}^{\rm SM}_{B}=\frac{g^4}{64\pi^2 M^2_W} \xi_t 
\left(\frac{F^{\rm SM}_{B}}{4}\right) (\bar\ell\ell)_{V-A} (\bar 
b d)_{V-A},
\end{eqnarray}
where
\begin{eqnarray}
\label{smbf}
F^{\rm SM}_{B}=\Big[\frac{x_t}{1-x_t}+\frac{x_t}{(1-x_t)^2} \ln x_t\Big].
\end{eqnarray}

Combining the penguin and box contributions the total amplitude can 
be written as,
\begin{eqnarray}
\label{sma}
\mathcal{M}^{\rm SM}=\frac{g^4}{64\pi^2 M_W^2} \xi_t F^{\rm SM}
(\bar\ell\ell)_{V-A}(\bar b d)_{V-A},
\end{eqnarray}
where
\begin{eqnarray}
\label{smf}
F^{SM}=\left[
\frac{F^{\rm SM}_B}{4} - \frac{F^{\rm SM}_{P}}{2}\right] = 
-\frac{x_t}{8}\left[\frac{x_t-4}{x_t-1}+\frac{3x_t}{(x_t-1)^2}\ln 
x_t\right].
\end{eqnarray}

A discussion on why the penguin contribution is $(V-A)(V-A)$ type is now
in order. The effective $Z\bar{b}d$ vertex has a pure $(V-A)$ form,
while an effective $\gamma\bar{b}d$ vertex vanishes identically. So a
photon cannot mediate the $B_d \to \ell^+\ell^-$ amplitude while a $Z$
boson can. Furthermore, since one can write
$J^{\mu}_{Z}=J^{\mu}_{3}-\sin^2\theta_W J^{\mu}_{\rm e.m.}$, only the
$t_3$ dependent part of the $Z$ coupling to final state leptons will
contribute to the decay amplitude, while the coefficient of $\sin^2
\theta_W$ in the amplitude would vanish identically.

Eq.~(\ref{sma}) leads to the following expression for the decay width
\cite{bb2}:
\begin{eqnarray}
\label{dkw}
\Gamma(B_d\rightarrow\ell^+\ell^-) = 
\frac{G_F^2}{\pi}\Big(\frac{\alpha}{4\pi\sin^2\theta_W}\Big)^2
m_B m_{\ell}^2 F_{B_d}^2 \xi_t^2\Big[F^{\rm SM}\Big]^2,
\end{eqnarray}
where $F_{B_d}$ is the decay constant.

\section{Extra dimensional contributions to
  \boldmath{$B_d \rightarrow\ell^+\ell^-$}}
In the NUED model, the decay $B_d\rightarrow\ell^+\ell^-$ will again
proceed via penguin and box contributions (see Figure 1). First, we
consider the penguin graphs.  Each such diagram will involve the KK
towers of $W^\pm$ or $\phi^\pm$ as internal propagators (inside the
loop), as well as the KK towers of $Z$ boson or a photon as the
external propagator (outside the loop) which couple to the final
leptons. If we take the zero modes of both the internal and external
lines, we get back the SM. Denoting the external propagator as $V$,
one can write the effective $V\bar b d$ vertex as
\begin{eqnarray}
\label{edevz}
\Gamma_\mu^{V} = -\frac{ig}{\cos\theta_W}\bigg(\frac{g^2}{16\pi^2}
F^{V}_{P}\xi_t\bigg)\gamma_\mu P_L,
\end{eqnarray}
where $V$ can be a zero mode $Z$ boson ($Z_{(0)}$), or a KK $Z$ boson
($Z_{(m)}$), or a KK photon ($\gamma_{(m)}$) (for the latter two cases
$m \neq 0$). As we have mentioned in the context of SM calculation,
the zero mode photon does not have any flavour changing effective
vertex. But this is not true for a KK photon -- we will discuss the
reasons shortly.  The $F^V_P$ functions in Eq.~(\ref{edevz}) capture
the KK loop dynamics when the indices of the internal
KK propagators are summed over. They take appropriate forms
$F_{Z_{(0)}}$, $F_{Z_{(m)}}$, or $(F_{\gamma_{(m)}}$ $\sin\theta_W
\cos\theta_W)$, depending on the nature of the external propagator
$V$.  In terms of these $F$ functions the penguin amplitude in the
NUED model can be written as
\begin{eqnarray}
\label{edpa}
\mathcal{M}^{\rm NUED}_{P} = \frac{g^4}{64\pi^2 M_W^2} \xi_t 
(\bar b d)_{V-A} \left[ \left(F_{Z_{(0)}} + F_{Z_{(m)}} \right)  
t_{3\ell} (\bar\ell\ell)_{V-A} - 2Q_\ell \sin^2\theta_W 
\left(F_{Z_{(m)}} - F_{\gamma_{(m)}} \right) (\bar\ell\ell)_V  
\right]. 
\end{eqnarray}
We now discuss one by one the KK dynamics of the three $F$ functions.

1. {\em $Z_{(0)}$ mediated penguins}:~ For each KK mode of the $W^\pm
   (\phi^\pm)$ propagator, the observations are exactly similar to the
   case of the SM penguins. The ultraviolet divergence cancels mode by
   mode when all the graphs for the same KK mode are added up. In
   terms of a new variable $x_n=1+({n^2\pi^2}/{a^2})$, one can write
   the $F$ function as
\begin{eqnarray}
\label{z0fz}
F_{Z_{(0)}}  
= 2\sum^{n_s}_{n=1}\frac{x_t}{4}\Big[\frac{6-x_t}{x_n-x_t} + 
\frac{x_n(-x_t+2)+4x_t}{(x_n-x_t)^2}\ln 
x_t +\frac{x_n(x_t-2)-4x_t}{(x_n-x_t)^2}\ln 
x_n\Big].
\end{eqnarray}
It should be noted that for $n = 0$ the above function reduces to 
$F^{\rm SM}_{P}$ as expected. Since here we are interested to quantify
the new physics part, we have not included  $n = 0$ case in the
summation. The sum is finite since for large $n$ each term in the sum
falls like $(\ln n/n^2)$. Clearly, the lighter KK modes dominate the sum.

2. {\em $Z_{(m)}$ mediated penguins}:~ Here the observations are very
   different from the SM case and there are quite a few issues that
   require attention. First of all, since the KK index of the external
   $Z$ propagator is nonzero, the contributions from both the zero and
   nonzero modes of the $W^\pm (\phi^\pm)$ loop propagators will have
   to be included within $F_{Z_{(m)}}$. Second, the $Z W W$, $Z W
   \phi$, and $Z \phi \phi$ vertices, which will appear in some of the
   penguin diagrams, do conserve KK parity. Hence, in such a diagram,
   say the one which involve $W^\pm$, instead of having two
   independent KK sums over the internal $W$ indices plus another over
   the KK modes of the external $Z$, there will be altogether two (and
   not three) independent KK sums. This makes that particular diagram
   less divergent than what it would have been without the bulk KK
   number conservation.

Next comes the question whether the net divergence cancels or
not. Curiously enough, here we encounter two sources of ultraviolet
divergences. The first one is the conventional source which we
encounter in the SM penguins too. From this perspective, as always,
each penguin diagram is ultraviolet divergent. In the SM, for exmple,
each penguin graph is logarithmically divergent. But in the present
case, the degree of divergence is different for different diagrams. In
this case, unlike in the SM, or for that matter in the case of
$Z_{(0)}$ mediated penguins with KK loop propagators, when we add up
all the amplitudes, the ultraviolet divergence does not mutually
cancel and exhibit a $(\ln n_s)^2$ kind of cut-off dependence.  The
second source of ultraviolet divergence arises when the KK summations
are carried over the so-called `finite' part (i.e. ultraviolet finite
for individual modes) for each diagram. Since each summation goes up
to $n_s$, what finally emerges after adding up the KK-summed otherwise
finite pieces from each diagram is a linearly divergent quantity
(i.e. proportional to $n_s$). Obviously, this second source
numerically dominates over the first one. The exact form of the $F$
function in this case is not easily tractable in a compact elegant
form (see Appendix A for individual expressions corresponding to each
diagram). For all practical purposes, it can be expressed as
\begin{eqnarray}
\label{zmfz}
F_{Z_{(m)}} \sim  a^2 n_s/\pi^2.
\end{eqnarray}
To understand the technical origin of this divergence, let us
concentrate on some specific diagrams. Take any penguin amplitude in
which the loop contains {\em two} bosons ($WW$, $W\phi$, or $\phi
\phi$). For $Z_{(m)}$ mediated case the KK indices of the two internal
bosons are independent; while, indeed, the index $m$ of $Z_{(m)}$ is
fixed by those two indices as a consequence of bulk KK number
conservation. So there will be {\em more than one} KK sum (in this
case, two) inside a one loop integral. This particular class of
diagrams is finally responsible for the non-cancellation of net
divergence. On the contrary, for $Z_{(0)}$ mediated amplitude,
whatever be the internal lines inside the loop, there is always a
single KK sum inside the integral, and the amplitude is
convergent. This is exactly what we have seen just before.

3. {\em $\gamma_{(m)}$ mediated penguins}:~ A KK photon behaves
   differently from the standard (zero mode) photon in the sense that
   while the latter does not have any off-diagonal fermionic couplings,
   the former has. This is again due to the fact that the KK photonic
   penguin amplitudes which contain the three-boson bulk vertices do
   contain a double summation over the KK indices of the propagators
   involving the loop momentum.  The $F$ function can be expressed as 
 \begin{eqnarray}
\label{npmfg}
F_{\gamma_{(m)}} \sim a^2 n_s/\pi^2.
\end{eqnarray}
Indeed, there is a order one difference between the coefficients of 
$F_{Z_{(m)}}$ and $F_{\gamma_{(m)}}$ which we do not display here.

Now we turn our attention to the box contribution involving KK towers of
$W^\pm$ and $\phi^\pm$. The box amplitude can be written as 
\begin{eqnarray}
\label{edab}
\mathcal{M}^{\rm NUED}_{B} = \frac{g^4}{64\pi^2 M_W^2} \sum_i \xi_i
\Big[\frac{F^{\rm NUED}_{B}}{4}\Big] (\bar bd)_{V-A}(\bar\ell\ell)_{V-A},
\end{eqnarray}
where
\begin{eqnarray}
\label{edfb}
F^{\rm NUED}_{B} & = & \sum_{n\neq 0}\sum_{m \neq 0}
\Big[\Big(\frac{2x_n\ln x_n-x_n}{2(x_n-x_i)}-\frac{2x_m\ln x_m-x_m}
{2(x_m-x_i)}\Big)
\frac{1}{x_n-x_m}+\frac{2x_i\ln x_i-x_i}{2(x_n-x_i)(x_m-x_i)}\Big]
\nonumber\\ 
&\sim&\ln n_s,
\end{eqnarray}
where $i$ runs over $u, c, t$ quarks. The simultaneous occurence of
zero modes is not included in the sum as that would correspond to the
SM contribution which we take separately.

Combining the KK-summed penguin and box graphs together, one can
calculate the relative change in the branching ratio 
${\cal{B}} (B_d \to \ell^+\ell^-)$ as 
\begin{eqnarray} 
\label{br}
\delta {\cal{B}}/{\cal{B}} = ({\cal{B}}^{\rm tot} -  
{\cal{B}}^{\rm SM})/{\cal{B}}^{\rm SM} \sim a^4 n_s^2/\pi^4. 
\end{eqnarray}

What about the allowed choices of $R$ and $n_s$? Measurements of Fermi
constant and electroweak precision observables put a lower limit on
$R^{-1}$ to be order 1 TeV \cite{ew}. $n_s$ should be restricted from
two perspectives. First of all, $n_s$ should not be too high such that
$M_s = n_s/R$ gets very close to the scale where gravity becomes
strong. Second, higher order loops in general would grow with higher
powers of $n_s$. For example, we have checked that the most divergent
two loop diagram for $B_d \to \ell^+\ell^-$ amplitude goes as
$a^2n_s^2 (\ln n_s)^2$. Calculation of the total two loop amplitude is
beyond the scope of the present analysis. The only thing we can say is
that $n_s$ should not be so large that the enhancement due to powers
of $n_s$ at a given order overshoots the usual loop suppression at
that order. Moreover, one must push up $R^{-1}$ to a rather large
value so that the scale ($M_s$) at which calculability is lost remains
sufficiently high, yet reasonably below the scale of strong gravity.

\section{UED versus NUED scenarios}
In this section we compare and contrast the renormalizability property
of the UED and NUED models with only one extra dimension. In the
former, all particles travel in the extra dimension, so the momentum
along the extra coordinate is always conserved. Therefore, the KK
number is conserved at every interaction point. Hence all the internal
propagator lines inside a one loop diagram have the {\em same} KK
number, hence a single summation, and the final amplitude is
well-behaved with $R$ as the only new physics parameter. On the other
hand, in the NUED scenario, the KK number is not conserved at the
brane though it is conserved in the bulk. The one loop propagators may
involve more than one KK indices, hence a multiple KK summation inside
the loop intergrals, and the final amplitude is ultraviolet
divergent. This is the key issue behind the nonrenormalizability of
the NUED model. The new physics parameters are both $R$ and $n_s$.

In the UED model there are some additional interactions which are not
present in the NUED model. Consider the scalars $W^5_{(n)}$, for
example. Can they couple to standard zero mode fermions? They do, but
in that case the other has to be a KK fermion in the same
representation of the zero mode fermion but with a wrong chirality.
The scalars $W^5_{(n)}$, $Z^5_{(n)}$ are in the adjoint representation
of the gauge group and do not have any zero mode.  It has been
discussed in \cite{buras1,buras2,debrupa} how such scalars mix with
the KK components of the doublet scalars to form three additional
physical scalars $a^0_{(n)}$ and $a^\pm_{(n)}$. These additional
scalars do not have any SM analog but their effects will have to be
included in the loop calculations in the UED model.  On the contrary,
in the NUED model $W^5_{(n)}$ or $Z^5_{(n)}$ cannot couple to the
brane localised SM fermions.

To provide more intuition into the impact of KK number conservation or
non-conservation, we consider some examples, which will illustrate how
the degree of divergence varies from process to process. We restrict
our comparison only up to one loop order.

1. While we have analysed in detail why $B_d \to \ell^+\ell^-$
amplitude is linearly divergent in NUED models, the same decay
amplitude is finite in UED models \cite{buras1}.

2. Consider the $B_d$--$\bar{B_d}$ or $K_0$--$\bar{K_0}$ mixing in the
NUED model \cite{santa1}. There are no bulk vertices, and the KK sums
for the two boson propagators can be carried out independently. Each
modified propagator increases the degree of divergence, and the final
amplitude is log divergent. On the other hand, if one does this
calculation in the UED model, the KK number being conserved at all
vertices, there is only a single KK summation.  Consequently, the final
result is ultraviolet finite \cite{santa2}. 

3. Consider the decay $Z \to b\bar{b}$. Since the external particles
are all zero mode states, the loop propagators will involve a single
KK index to be summed regardless of the UED or NUED nature of the
model. In either case, the result is $n_s$ independent
\cite{santa1,santa2}. 

4. We have calculated the amplitude $t \to c Z$ in the NUED model.  As
expected, we obtain a cut-off independent amplitude since there is
only a single KK summation. The numerical modifications are not
significant. 

5. In view of the arguments advanced above, it is not difficult to see
that $b \to s \gamma$ amplitude is finite, but $b \to s \ell^+\ell^-$
amplitude is linearly divergent in NUED models. Both are finite in UED
models \cite{buras2}.

\section{Conclusions} 
The extra dimensional models are all nonrenormalizable due to the
multiplicity of KK states. We know that one loop wavefunction
renormalizations are linearly divergent leading to power law running
of the couplings \cite{powerlaw}. But when it comes to the calculation
of physical observables, in some special cases there may be a
cancellation between wavefunction renormalizations and vertex
corrections leading to finite results (only at one loop). Our aim in
this paper has been to identify the root cause of
nonrenormalizability, in particular, the degree of divergence in
different physical observables both in UED and NUED models. The
crucial question is whether KK number is conserved or not.  A summary
of the situation is the following:

1. In the NUED model with only one extra dimension the tree level KK
   summed amplitude is finite. In the UED picture at the tree level
   there is no KK summation. 

2. Next consider one loop corrections with one extra dimension. This
constitutes the main thesis of our paper.  While in the UED scenario
the amplitude is finite, the NUED scenario suffers from ultraviolet
sensitivity. In the latter framework, the $B_d \to \ell^+\ell^-$
amplitude provides a nice platform to examine the technical origin of
the cut-off dependence as the KK indices are contained both in the
`internal' (i.e., inside the loop integral) and `external'
propagators. The interplay between the `internal' and `external' KK
indices while being summed, and the number of such independent
summations inside the loops, count to determine the degree of
ultraviolet divergence. In the NUED scenario, (i) $B_d \to
\ell^+\ell^-$ and $b \to s \ell^+\ell^-$ amplitudes are linearly
divergent, (ii) $B$--$\bar{B}$ or $K$--$\bar{K}$ mixing is
logarithmically divergent, while (iii) $Z \to b\bar{b}$, $t \to c Z$
and $b \to s \gamma$ decay amplitudes are finite. With the basic
observations outlined in the previous sections, one can look at other
processes and check their ultraviolet sensitivities.

3. Now consider two (or higher) loop corrections, again with only one
   extra dimension. The UED results are now divergent, while NUED
   results would be {\em more} divergent because the latter involves
   more KK summations. 

4. Finally, take more than one extra dimension. Both UED and NUED
   models give divergent results at any order. As expected, the latter
   is more divergent than the former.   

\section*{Acknowledgements}
Both of us acknowledge hospitality of the Abdus Salam ICTP, Trieste,
while a significant part of the work was being done. We thank E. Dudas
and A. Raychaudhuri for reading the manuscript and valuable
suggestions. We also thank T. Aliev and P.B. Pal for discussions.
G.B.'s research has been supported, in part, by the DST, India,
project number SP/S2/K-10/2001.

\vskip 5pt
\catcode`@=11 \@addtoreset{equation}{section} \catcode`@=12
\renewcommand{\theequation}{\thesection.\arabic{equation}}

\appendix{}
\vskip -20pt
\section{Relevant loop integrals in the SM and NUED model}

For the $Z$-penguin diagrams, the expressions for the 
effective $Z\bar{b}d$ vertex for the individual diagrams are 
given below (for each expression the loop propagators are written
within square bracket).
\begin{eqnarray}
\label{list1}
\Gamma^{\mu} [Wq_iq_i]&=
&-\frac{ig}{\cos\theta_W}\bigg(\frac{g^2}{16\pi^2}
\sum_{i}\xi_i\Big[2a^i_LI_{3}+a^i_Rx_iI_{1}\Big]\bigg)\gamma_\mu
P_L\nonumber ,\\ 
\Gamma^{\mu} [\phi
q_iq_i]&=
&-\frac{ig}{\cos\theta_W}\bigg(\frac{g^2}{16\pi^2}
\sum_{i}\xi_i\frac{x_i}{2}\Big[a^i_R(2I_{3}-1/2)+a^i_Lx_iI_{1}\Big]\bigg)
\gamma_{\mu}P_L\nonumber ,\\
\Gamma^{\mu} [W W
q_i]&=&-\frac{ig}{\cos\theta_W}\bigg(\frac{g^2}{16\pi^2}
\sum_{i}\xi_i\Big[-6\cos^2\theta_W
I_{4}\Big]\bigg)\gamma_{\mu}P_L\nonumber ,\\
\Gamma^{\mu} [\phi\phi
q_i]&=
&-\frac{ig}{\cos\theta_W}\bigg(\frac{g^2}{16\pi^2}
\sum_{i}\xi_i\Big[\cos2\theta_W\frac{x_i}{2}I_{4}\Big]\bigg)
\gamma_{\mu}P_L , \\
\Gamma^{\mu} [W\phi
q_i]&=
&-\frac{ig}{\cos\theta_W}\bigg(\frac{g^2}{16\pi^2}
\sum_{i}\xi_i\Big[\sin^2\theta_W x_i I_{2}\Big]\bigg)
\gamma_{\mu}P_L\nonumber ,\\
\Gamma^{\mu} [W q_i]&=
&-\frac{ig}{\cos\theta_W}\bigg(\frac{g^2}{16\pi^2}
\sum_{i}\xi_i\Big[-a^d_LI_{5}\Big]\bigg)\gamma_{\mu}P_L\nonumber ,\\
\Gamma^{\mu} [\phi
q_i]&=
&-\frac{ig}{\cos\theta_W}\bigg(\frac{g^2}{16\pi^2}
\sum_{i}\xi_i\Big[-a^d_L\frac{x_i}{2}I_{5}\Big]\bigg)\gamma_{\mu}P_L
\nonumber .
\end{eqnarray}
For the zero mode $Z$ propagators the $I_i$ functions are given by, 
\begin{eqnarray}
\label{list3}
\int\frac{d^4k}{(2\pi)^4}2
\sum_{n=1}^{n_s}\frac{1}{(k^2-m^2_i)^2(k^2-M^2_{W_n})}&=
&-\frac{i}{16\pi^2M^2_W}I^{Z_{(0)}}_{1}\nonumber ,\\
\int\frac{d^4k}{(2\pi)^4}2
\sum_{n=1}^{n_s}\frac{k^{\mu}k^{\nu}}{(k^2-m^2_i)^2(k^2-M^2_{W_n})}&=
&\frac{ig^{\mu\nu}}{16\pi^2}I^{Z_{(0)}}_{3}\nonumber ,\\
\int\frac{d^4k}{(2\pi)^4}2
\sum_{n=1}^{n_s}\frac{k^{\mu}}{(k^2-m^2_i)[(k-p)^2-M^2_{W_n}]}&=
&\frac{ip^{\mu}}{16\pi^2}I^{Z_{(0)}}_{5} ,\\
\int\frac{d^4k}{(2\pi)^4}2
\sum_{n=1}^{n_s}\frac{1}{(k^2-m^2_i)(k^2-M^2_{W_n})^2}&=
&-\frac{i}{16\pi^2M^2_W}I^{Z_{(0)}}_{2}\nonumber ,\\
\int\frac{d^4k}{(2\pi)^4}2
\sum_{n=1}^{n_s}\frac{k^{\mu}k^{\nu}}{(k^2-m^2_i)(k^2-M^2_{W_n})^2}&=
&\frac{ig^{\mu\nu}}{16\pi^2}I^{Z_{(0)}}_{4} \nonumber .
\end{eqnarray}
Above, we have used $M_{W_n}^2=M_W^2+(n/R)^2$. 
Since we take the SM contributions separately, to avoid any
double-counting, we carry out the sum over the KK index $n$
from $1$ to $n_s$ and multiply by an overall factor of 2,
instead of performing the sum from $-n_s$ to $+n_s$.

When the $Z$ propagator has nonzero KK mode, the propagator itself is 
modified as,
\begin{eqnarray}
\label{list4}
\frac{1}{M_Z^2}\longrightarrow2\sum_{m=1}^{n_s}\frac{1}{M_Z^2+n^2/R^2}\sim
\frac{1}{M_Z^2}\bigg[2\sum_{m=1}^{n_s}\frac{a^2}{A\pi^2m^2}\bigg],
\end{eqnarray}
where $A=(M_W/M_Z)^2$. The loop integrals are also modified, and the 
corresponding expressions for the $I_i$
are given by, 
\begin{eqnarray}
\label{list5}
\int\frac{d^4k}{(2\pi)^4}2
\sum_{m=1}^{n_s}\frac{a^2}{A\pi^2m^2}
\sum_{n=-n_s}^{n_s}\frac{1}{(k^2-m^2_i)^2(k^2-M^2_{W_n})}&=
&-\frac{i}{16\pi^2M^2_W}I^{Z_{(m)}}_{1}\nonumber ,\\
\int\frac{d^4k}{(2\pi)^4}2\sum_{m=1}^{n_s}\frac{a^2}{A\pi^2m^2}
\sum_{n=-n_s}^{n_s}\frac{k^{\mu}k^{\nu}}{(k^2-m^2_i)^2(k^2-M^2_{W_n})}&=
&\frac{ig^{\mu\nu}}{16\pi^2}I^{Z_{(m)}}_{3}\nonumber ,\\
\int\frac{d^4k}{(2\pi)^4}2\sum_{m=1}^{n_s}\frac{a^2}{A\pi^2m^2}
\sum_{n=-n_s}^{n_s}\frac{k^{\mu}}{(k^2-m^2_i)[(k-p)^2-M^2_{W_n}]}&=
&\frac{ip^{\mu}}{16\pi^2}I^{Z_{(m)}}_{5} ,\\
\int\frac{d^4k}{(2\pi)^4}2\sum_{m=1}^{n_s}\frac{a^2}{A\pi^2m^2}
\sum_{n=-n_s+m}^{n_s}\frac{1}{(k^2-m^2_i)(k^2-M^2_{W_n})
(k^2-M^2_{W_{(n-m)}})}&=
&-\frac{i}{16\pi^2M^2_W}I^{Z_{(m)}}_{2}\nonumber ,\\
\int\frac{d^4k}{(2\pi)^4}2\sum_{m=1}^{n_s}\frac{a^2}{A\pi^2m^2}
\sum_{n=-n_s+m}^{n_s}\frac{k^{\mu}k^{\nu}}{(k^2-m^2_i)(k^2-M^2_{W_n})
(k^2-M^2_{W_{(n-m)}})}&=&\frac{ig^{\mu\nu}}{16\pi^2}I^{Z_{(m)}}_{4}
\nonumber.
\end{eqnarray}
Note that the limits of summation in the last two relations are so
chosen that none of the KK numbers exceeds $n_s$. 
Also notice that in the last two integrals despite the presence of
three KK propagators, KK number conservation ensures only two
independent summation. This effectively reduces the degree of
ultraviolet divergence. 

Clearly, when all the KK indices are zero, the integrals correspond to
the SM case.

\newpage

\noindent
\begin{figure}[htbp]
\begin{center}\begin{picture}(450,350)(0,0)
\ArrowLine(50,300)(10,300) \Text(30,308)[]{$b$}
\ArrowLine(10,240)(50,240) \Text(30,235)[]{$d$}
\DashLine(50,240)(50,300){4} \Text(32,270)[]{$W,\phi$}
\DashLine(48,240)(48,300){4}
\ArrowLine(100,272)(50,300)\Text(79,297)[]{$q_i$}
\ArrowLine(50,240)(100,268)\Text(79,245)[]{$q_i$}
\Photon(100,272)(140,272){4}{4}\Text(120,288)[]{$Z$}
\Photon(100,268)(140,268){4}{4}
\ArrowLine(140,272)(180,300) \Text(160,297)[]{$\ell$}
\ArrowLine(180,240)(140,268) \Text(160,245)[]{$\ell$}
\Vertex(49,300){2.0} \Vertex(49,240){2.0}\Vertex(100,270){2.0}
\Vertex(140,270){2.0}

\ArrowLine(300,300)(260,300) \Text(280,308)[]{$b$}
\ArrowLine(260,240)(300,240) \Text(280,235)[]{$d$}
\ArrowLine(300,240)(300,300) \Text(290,270)[]{$q_i$}
\DashLine(350,272)(300,300){4}\Text(329,297)[]{$W,\phi$}
\DashLine(350,270)(300,298){4}
\DashLine(300,240)(350,268){4}\Text(329,245)[]{$W,\phi$}
\DashLine(300,242)(350,270){4}
\Photon(350,272)(390,272){4}{4}\Text(370,288)[]{$Z$}
\Photon(350,268)(390,268){4}{4}
\ArrowLine(390,272)(430,300) \Text(410,297)[]{$\ell$}
\ArrowLine(430,240)(390,268) \Text(410,245)[]{$\ell$}
\Vertex(299,300){2.0} \Vertex(299,240){2.0}\Vertex(350,270){2.0}
\Vertex(390,270){2.0}

\ArrowLine(10,150)(50,150) \Text(30,158)[]{$\ell$}
\ArrowLine(50,150)(180,150) \Text(115,158)[]{$\ell$}
\ArrowLine(10,90)(50,90) \Text(30,82)[]{$d$}
\ArrowLine(50,90)(90,90) \Text(70,82)[]{$d$}
\ArrowLine(85,90)(140,90) \Text(110,82)[]{$q_i$}
\ArrowLine(140,90)(180,90) \Text(160,82)[]{$b$}
\Photon(49,150)(49,90){4}{4}\Text(38,120)[]{$Z$}
\Photon(51,150)(51,90){4}{4}
\DashCArc(110,90)(25,0,180){4}\Text(110,125)[]{$W,\phi$}
\DashCArc(110,90)(23,0,180){4}
\Vertex(50,150){2.0} \Vertex(49,90){2.0}\Vertex(85,90){2.0}
\Vertex(135,90){2.0}

\ArrowLine(260,150)(390,150) \Text(325,158)[]{$\ell$}
\ArrowLine(390,150)(430,150) \Text(410,158)[]{$\ell$}
\ArrowLine(260,90)(300,90) \Text(280,82)[]{$d$}
\ArrowLine(294,90)(350,90) \Text(325,82)[]{$q_i$}
\ArrowLine(350,90)(390,90) \Text(370,82)[]{$b$}
\ArrowLine(390,90)(430,90) \Text(410,82)[]{$b$}
\Photon(391,150)(391,90){4}{4}\Text(402,120)[]{$Z$}
\Photon(389,150)(389,90){4}{4}
\DashCArc(320,90)(25,0,180){4}\Text(319,125)[]{$W,\phi$}
\DashCArc(320,90)(23,0,180){4}
\Vertex(390,150){2.0} \Vertex(296,90){2.0}\Vertex(344,90){2.0} 
\Vertex(390,90){2.0}

\ArrowLine(175,0)(135,0) \Text(155,8)[]{$b$}
\ArrowLine(135,-60)(175,-60) \Text(155,-65)[]{$d$}
\ArrowLine(175,-60)(175,0) \Text(167,-30)[]{$q_i$}
\DashLine(175,0)(265,0){4}\Text(220,13)[]{$W,\phi$}
\DashLine(175,-2)(265,-2){4}
\ArrowLine(305,0)(265,0) \Text(285,8)[]{$\ell$}
\ArrowLine(265,-60)(305,-60) \Text(285,-68)[]{$\ell$}
\ArrowLine(265,0)(265,-60) \Text(273,-30)[]{$\nu_{\ell}$}
\DashLine(175,-60)(265,-60){4}\Text(220,-70)[]{$W,\phi$}
\DashLine(175,-58)(265,-58){4}
\Vertex(175,0){2.0} \Vertex(175,-60){2.0} \Vertex(265,0){2.0} 
\Vertex(265,-60){2.0}

\end{picture}
\end{center}
\vskip 100pt
\caption[]{\small\sf The Feynman graphs contributing to $B_d \to
  \ell^+\ell^-$. The bosons are shown as double lines (dashed and wavy)
  which can have zero or non-zero KK modes.}

\end{figure}
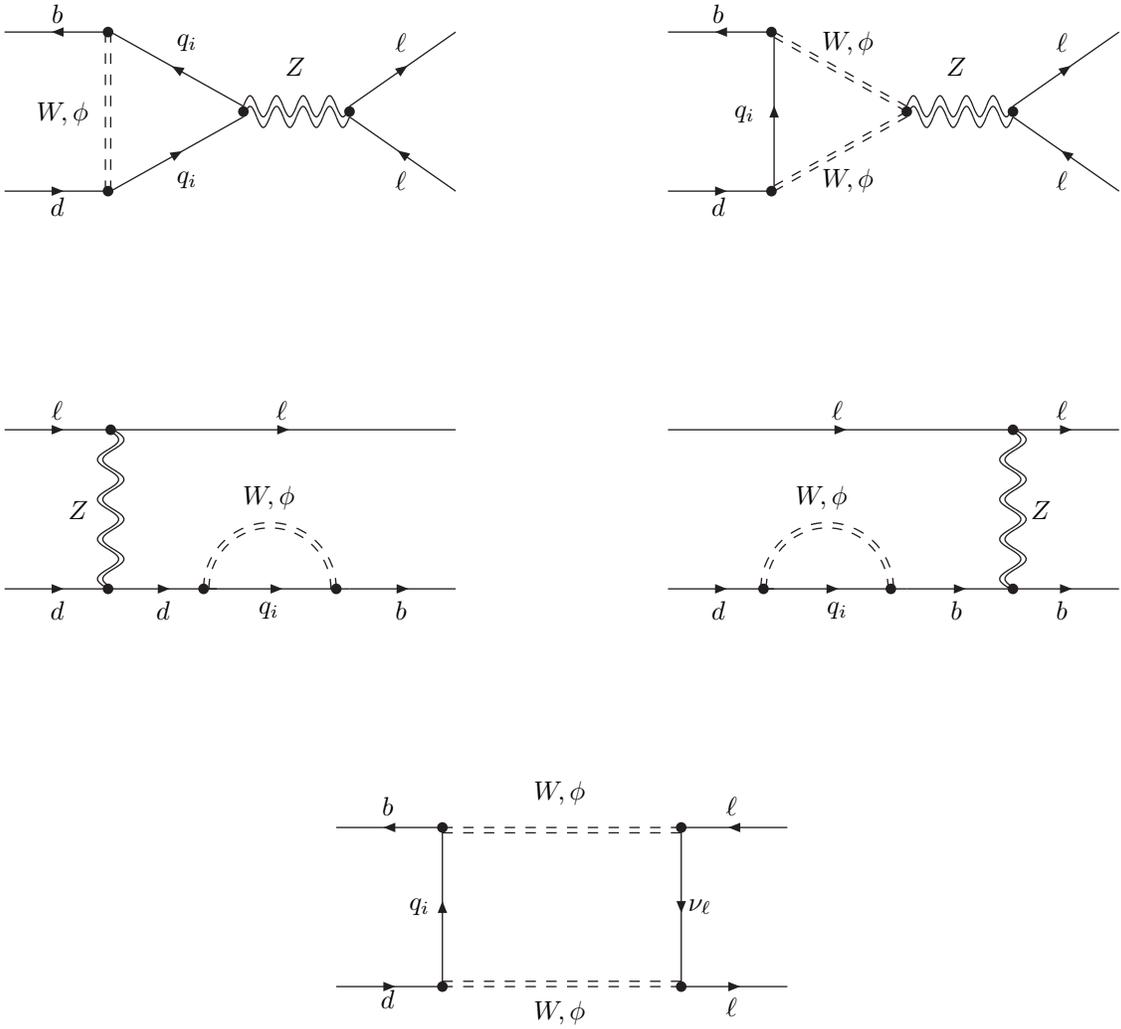

\end{document}